\documentclass[aps,nofootinbib,reprint,endfloats,showpacs,showkeys]{revtex4-1}
\pdfoutput=1

\usepackage{hyperref}
\usepackage{graphicx}
\usepackage{epsfig}
\usepackage{amsmath}

\begin{document}

\title{Transport Equations for Oscillating Neutrinos}

\author{Yunfan Zhang}
 \altaffiliation[Now at ]{Scripps Institution of Oceanography, University of California, San Diego, La Jolla, CA 92093-0208 USA.}
\affiliation{Department of Physics\\ Princeton University\\ Princeton, NJ 08544 USA
\\ e-mail: yf.g.zhang@gmail.com}

\author{Adam Burrows}
\affiliation{Department of Astrophysical Sciences\\ Princeton University\\ Princeton, NJ 08544 USA
\\ e-mail: burrows@astro.princeton.edu}  %%%%\\ URL: \url{http://www.astro.princeton.edu/~burrows}}
 \homepage{http://www.astro.princeton.edu/~burrows}

%\email[]{burrows@astro.princeton.edu}
%\homepage[]{http://www.astro.princeton.edu/~burrows}
%\thanks{}
%\altaffiliation{}

\begin{abstract}
We derive a suite of generalized Boltzmann equations, based on the density-matrix
formalism, that incorporates the physics of neutrino oscillations for two- 
and three-flavor oscillations, matter refraction, and self-refraction.  The resulting 
equations are straightforward extensions of the classical transport equations that nevertheless
contain the full physics of quantum oscillation phenomena.  In this way,  our broadened 
formalism provides a bridge between the familiar neutrino transport algorithms employed by supernova modelers 
and the more quantum-heavy approaches frequently employed to illuminate the various neutrino oscillation effects.
We also provide the corresponding angular-moment versions of this generalized equation set.
Our goal is to make it easier for astrophysicists to address oscillation phenomena in a language with which
they are familiar.  The equations we derive are simple and practical, and are intended to 
facilitate progress concerning oscillation phenomena in the context of core-collapse supernova theory.
\end{abstract}



\maketitle

\onecolumngrid

\section{Introduction}
\label{intro}

Core-collapse supernova explosions, whatever their mechanism, involve
neutrino transport and neutrino-matter coupling in a fundamental way \citep{2007PhR...442...38J}.
At the extreme densities and temperatures encountered in the unstable
stellar core there is prodigious production of neutrinos of all six
neutrino species ($\nu_{e}$, $\bar{\nu}_{{e}}$, $\nu_{\mu}$, $\bar{\nu}_{{\mu}}$, $\nu_{\tau}$, and $\bar{\nu}_{{\tau}}$)
and the corresponding integrated neutrino flux can be comparable to
the dynamical matter flux.  The consensus mechanism of explosion centrally
involves neutrino heating of the shocked mantle to drive the blast to infinity,
leaving behind a cooling and deleptonizing proto-neutron star \citep{2013RvMP...85..245B}.  Hence, to
understand these phenemona in any way requires an understanding of
neutrino physics and transport in every particular.

The mathematical description of neutrino transport and transfer frequently
starts with a classical Boltzmann equation for the corresponding phase-space density ($f_{\nu}$)
for each species \citep{2007ApJ...659.1458H}.  An equivalent formulation would solve for the specific intensity ($I_{\nu}$)
of the multi-angle, multi-energy-group neutrino beams at every spatial point, at every time,
for every species \citep{Mihalas1999,Castor2004}.  The solution of this time-dependent, multi-angle, spectral field for all spacetime
and for six species is a numerical ``grand challenge" of the first order that remains out of
computational reach.  Nevertheless, Boltzmann equations are the starting points for
the various more simplified approaches to neutrino transport that to date have been employed
by supernova theorists.  These include multi-group, flux-limited diffusion and
two-moment closures \citep{2013ApJS..204....7Z,2011JQSRT.112.1323V}.  Moreover, to lessen the computational burden of the
challenging non-spherical hydrodynamic context revealed by current theory to be important,
researchers have oftimes reduced the transport problem even further into multiple one-dimensional radial/spherical
solves (the so-called ``ray-by-ray" method \citep{1995ApJ...450..830B}). Though less computationally demanding, this dimensional
compromise may be problematic \citep{2013RvMP...85..245B}.

However, we now know that neutrinos have mass and oscillate among themselves \citep{PhysRevD.58.096016,2013ApJ...765...14L}.  In addition,
neutrino-matter refraction can lead to resonant conversion between species, even if the vacuum
oscillation angles are small \citep{1978PhRvD..17.2369W,1979PhRvD..20.2634W,Langacker2010}.  Furthermore, it has been shown that neutrino-neutrino refraction
effects in the neutrino-rich supernova environment can lead to self-oscillation effects \citep{1992PhLB..287..128P,2006PhRvD..74l3004D,
2010ARNPS..60..569D} and in particular ``spectral swapping/splitting" 
\citep{2007PhRvL..99x1802D,2007PhRvD..76h5013D,2007PhRvD..75l5005D,2008PhRvD..77h5016D,PhysRevLett.103.051105,2007PhRvD..76h1301R,2012arXiv1209.5894B}.
These effects are most prominent in supernovae if the mass hierarchy \citep{2013arXiv1303.0758R,2013arXiv1302.0779D} 
is inverted (``IH", a possibility) or when (rarely) the electron number density is not greater than 
that of the neutrinos \citep{2011PhRvL.106i1101D,2012PhRvD..85k3007S,PhysRevD.85.085031}.  However, the possibility of such a rich set of
oscillation behaviors and transformations has introduced new excitement into supernova
science \citep{2012PhRvD..85f5008D,2012MNRAS.425.1083P,2011ApJ...738..165S}.  At the very least, oscillation phenomena will alter the mix and spectra of
supernova neutrinos detected at Earth \citep{PhysRevC.78.015807,2002astro.ph..5390S,2012ARNPS..62...81S,PhysRevD.85.085031}.

Neutrino oscillations are purely quantum-mechanical effects that are not captured by the classical Boltzmann equation.
However, as was shown by Strack \& Burrows \citep{2005PhRvD..71i3004S} and Sigl \& Raffelt \citep{1993NuPhB.406..423S,1993PhRvL..70.2363R}, the density matrix formalism
of quantum mechanics is the most natural generalization of classical transport theory (and its simplifications)
with which to incorporate oscillation physics.  The angle- and energy-dependent effects of self-refraction
are naturally accounted for.  Whereas the classical formulation involves only the diagonal components
of the density matrix (each component being associated with a given neutrino species), the quantum-mechanical
extension adds corresponding equations for the off-diagonal ``phase-space densities," and new sources
on the ``right-hand-sides" that now couple the various neutrino species. These sources are added
to the classical absorption, scattering, and emission sources of classical transport. From the
oscillation perspective, the latter classical terms represent the various physical decoherence
effects, and comparisons between the magnitudes of these terms and the oscillation coupling terms
provide a natural means to determine the potential degrees of decoherence.

In a real sense, the quantum-mechanical equation set merely increases the number of
partial differential equations similar to the classical Boltzmann equation that need to be simultaneously solved.
This makes it easier for the supernova theorist and/or numericist to generalize their 
classically developed formalism to include neutrino oscillations in a quantum-mechanically rigorous way.
There are no wave functions, and in principle no imaginary quantities need be addressed nor invoked.
This provides a much-needed bridge between the oftimes opaque oscillation literature and the practical
astrophysicist. ``All" that needs to be done is to solve an extended set of coupled Boltzmann-like equations.

It is with this philosophy in mind that we present in this paper the generalized set of neutrino transport
equations that contain oscillation physics.  We do this for both two-neutrino and three-neutrino variants.
We also provide the associated moment equations that might lead to practical and tractable simplifications
of this equation set.  For the latter, we do not provide the higher-moment
closures necessary to employ the moment approach.  This art we leave to future work. However, we believe the
format of the resultant moment equations is particularly clear and should prove useful.  No attempt is made to address the
energy and angular resolutions that may be necessary to manifest all the various possible oscillation effects.
Nor do we suggest various averaging procedures over neutrino energy, etc. that might obviate the need for
high energy or angle resolution.  All approaches and studies of oscillation phenomena are burdened with the same
concerns and issues.  In this spirit, since the various consequences of neutrino oscillations and refraction have been
been amply explored in the literature, and Strack \& Burrows \citep{2005PhRvD..71i3004S} have already
demonstrated that the density-matrix approach yields the correct results for a subset of them (see also \citep{2012arXiv1202.2243B}),
we won't in this paper provide any solutions to the equations derived.  Rather, we hope that those in the supernova
community interested in generalizing their thinking and calculations to incorporate oscillation phenomena
with minimal effort will find our results of use.

\section{Transport Equation with Oscillations \label{sec:Transport-Equation-with}}

\subsection{Neutrino Mixing \label{sub:Neutrino-Mixing}}

Neutrinos exhibit mixing because of the discrepancy between flavor
and mass states. For two flavors, the transformation from the mass basis
to the flavor basis is given by 
\begin{equation}
\left(\begin{array}{c}
\nu_{1}\\
\nu_{2}
\end{array}\right)=U\left(\begin{array}{c}
\nu_{e}\\
\nu_{\mu}
\end{array}\right),
\end{equation}
where, with mixing angle $\theta$
\begin{equation}
U=\left(\begin{array}{cc}
\cos\theta & \sin\theta\\
-\sin\theta & \cos\theta
\end{array}\right).
\end{equation}

For three flavors, the transformation to the flavor basis is given by
the Pontecorvo-Maki-Nakagawa-Sakata (PMNS) matrix \citep{Langacker2010} $U$: 
\begin{equation}
\left(\begin{array}{c}
\nu_{1}\\
\nu_{2}\\
\nu_{3}
\end{array}\right)=U\left(\begin{array}{c}
\nu_{e}\\
\nu_{\mu}\\
\nu_{\tau}
\end{array}\right),
\end{equation}
where in this case, with $c_{ij}=\cos\theta_{ij}$, and $s_{ij}=\sin\theta_{ij}$
\begin{equation}
U=\left(\begin{array}{ccc}
1 & 0 & 0\\
0 & c_{23} & s_{23}\\
0 & -s_{23} & c_{23}
\end{array}\right)\left(\begin{array}{ccc}
c_{13} & 0 & s_{13}e^{-i\delta}\\
0 & 1 & 0\\
-s_{13}e^{i\delta} & 0 & c_{13}
\end{array}\right)\left(\begin{array}{ccc}
c_{12} & s_{12} & 0\\
-s_{12} & c_{12} & 0\\
0 & 0 & 1
\end{array}\right).\label{eq:024}
\end{equation}

An example of the three mixing angles and $\Delta m^{2}$ in the currently
accepted range is summarized in Table \ref{tab:Neutrino-Mixing-Parameters}
for both the normal hierarchy (NH) and the inverted hierarchy (IH)
\citep{2012JHEP...12..123G,2012PhRvD..86a3012F}. The CP phase $\delta$
has not been determined. 

\subsection{The Heisenberg-Boltzmann Equation \label{sub:The-Heisenberg-Boltzmann-Equatio}}

The Heisenberg Boltzmann Equation is given by \citep{2005PhRvD..71i3004S}

\begin{equation}
\frac{\partial\mathcal{F}}{\partial t}+\vec{v}\cdot\frac{\partial\mathcal{F}}{\partial\vec{r}}+\dot{\vec{p}}\cdot\frac{\partial\mathcal{F}}{\partial\vec{p}}=-i[H,\mathcal{F}]+C,\label{eq:5}
\end{equation}
where for three flavors of neutrinos %
\begin{equation}
\mathcal{F}=\langle\nu|\rho|\nu\rangle=\left(\begin{array}{ccc}
\langle\nu_{e}|\nu_{e}\rangle & \langle\nu_{e}|\nu_{\mu}\rangle & \langle\nu_{e}|\nu_{\tau}\rangle\\
\langle\nu_{\mu}|\nu_{e}\rangle & \langle\nu_{\mu}|\nu_{\mu}\rangle & \langle\nu_{\mu}|\nu_{\tau}\rangle\\
\langle\nu_{\tau}|\nu_{e}\rangle & \langle\nu_{\tau}|\nu_{\mu}\rangle & \langle\nu_{\tau}|\nu_{\tau}\rangle
\end{array}\right)\label{eq:6}
\end{equation}
 is the density matrix given by the Wigner phase space density
\begin{equation}
\rho(r,p,t)=\int d^{3}Re^{-ipR}\psi^{\dagger}(r-\frac{R}{2},t)\psi(r+\frac{R}{2},t)\, ,
\label{eq:7}
\end{equation}
where $p$ denotes the particle momentum.
$C$ denotes the classical scattering, absorption, and emission terms, and is given by:
\begin{equation}
C=\left(\begin{array}{ccc}
C_{\nu_{e}} & 0 & 0\\
0 & C_{\nu_{\mu}} & 0\\
0 & 0 & C_{\nu_{\tau}}
\end{array}\right)\ .\label{eq:8}
\end{equation}
Note the $C$ is diagonal and does not couple different neutrino species.
The analogous form of Eq. \eqref{eq:6} and \eqref{eq:8} for two flavors is 
straightforward \footnote{
For two flavors one has
\begin{equation}
\mathcal{F}=\langle\nu|\rho|\nu\rangle=\left(\begin{array}{cc}
\langle\nu_{e}|\nu_{e}\rangle & \langle\nu_{e}|\nu_{\mu}\rangle\\
\langle\nu_{\mu}|\nu_{e}\rangle & \langle\nu_{\mu}|\nu_{\mu}\rangle
\end{array}\right)\label{eq:9}
\end{equation}
and
\begin{equation}
C=\left(\begin{array}{cc}
C_{\nu_{e}} & 0\\
0 & C_{\nu_{\mu}}
\end{array}\right)\ .
\end{equation}
}
.
We note that the last term on the left-hand-side of Eq. \eqref{eq:5}
represents gravitational redshift, and can be omitted if general relativistic
effects on neutrino transport are not considered. 

Note that $\mathcal{F}$ is Hermitian, and can thus be expanded in unitary groups.
We define real quantities $f_{\gamma}$ so that
\begin{equation}
\mathcal{F}=\sum_{\gamma=0}^{3}f_{\gamma}\sigma^{\gamma}\label{eq:11}
\end{equation}
for two flavors and 
\begin{equation}
\mathcal{F}=\sum_{\gamma=0}^{8}f_{\gamma}\lambda^{\gamma}\label{eq:12}
\end{equation}
for three flavors, where $\sigma$ and $\lambda$ are the Dirac and
Gell-Mann matrices (see Appendix \ref{sec:Expansion}). Note that the $f_{\gamma}$
for two flavors are not related to those for three flavors.
This structure arises from the SU(N) rotation symmetry in neutrino flavor space that the transformations due to neutrino
oscillations must satisfy \citep{2005PhRvD..72d5003S,2011arXiv1111.2282B,2012arXiv1209.5894B}.
The Hamiltonian contains the terms 
\begin{equation}
H=H_{0}+H_{e}+H_{\nu\nu}-H_{\nu\bar{\nu}}^{*},
\end{equation}
where {*} denotes complex conjugate. 
For two flavors, the vacuum and matter Hamiltonians $H_{0}$ and $H_{e}$
in the flavor basis take the familiar form 
\begin{equation}
\begin{aligned}H_{0}^{1} & =\frac{\omega}{2}\sin2\theta\sigma^{1}-\frac{\omega}{2}\cos2\theta\sigma^{3},\end{aligned}
\end{equation}
and 
\begin{equation}
H_{e}=\frac{A}{2}\sigma^{3},
\end{equation}
where the vacuum frequency $\omega$ ($(m_1^2-m_2^2)/2p$) and mixing angle $\theta$ take
usual meanings, and $A=\frac{\sqrt{2}G_{F}n_{e}}{\hbar}$ represents
the interaction strength. $n_e$ is the electron number density.

For three flavors, the traceless vacuum Hamiltonian in the flavor basis
is 
\begin{equation}
H{}_{0}=\frac{1}{3}U\left(\begin{array}{ccc}
\frac{-\Delta_{21}-\Delta_{31}}{2p}\\
 & \frac{\Delta_{21}-\Delta_{32}}{2p}\\
 &  & \frac{\Delta_{32}+\Delta_{31}}{2p}\, ,
\end{array}\right)U^{\dagger}.\label{eq:0025}
\end{equation}
where $\Delta_{ij} = m_i^2 - m_j^2$ and $p$ is the particle momentum/energy.
From here the expression $H_{0}^{\gamma}$ in 
\begin{equation}
H_{0}=\sum_{\gamma=0}^{8}H_{0}^{\gamma}\lambda_{\gamma}
\end{equation}
can be extrapolated 
\footnote{For example, ignoring the CP phase, and denoting $\omega_{31}=\frac{\Delta_{31}}{2 p}$,
$\omega_{21}=\frac{\Delta_{21}}{2 p}$, $\omega_{32}=\omega_{31}-\omega_{21}$,
$S_{ij}=\sin2\theta_{ij}$, and $K_{ij}=\cos2\theta_{ij}$, we have
\begin{equation}
\begin{aligned}2H_{0}^{1} & =(\omega_{31}-\omega_{21}s_{12}^{2})s_{23}S_{13}+\omega_{21}c_{13}c_{23}S_{12},\\
2H_{0}^{4} & =(\omega_{31}-\omega_{21}s_{12}^{2})c_{23}S_{13}-\omega_{21}c_{13}s_{23}S_{12},\\
2H_{0}^{6} & =-\omega_{21}s_{13}S_{12}K_{23}+\frac{S_{23}}{4}\left[4\omega_{31}c_{13}^{2}-\omega_{21}(1+3K_{12}+2K_{13}s_{12}^{2})\right],\\
2H_{0}^{3} & =\frac{1}{8}\left[-(\omega_{21}-2\omega_{31})(1-3K_{13}+2K_{23}c_{13}^{2})+\omega_{21}\left(4S_{12}S_{23}s_{13}-K_{12}(6c_{13}^{2}+(3-K_{13})K_{23}\right)\right],\\
2H_{0}^{8} & =\frac{1}{24\sqrt{3}}\left[-(\omega_{21}-2\omega_{31})(1-3K_{13}-6K_{23}c_{13}^{2})-3\omega_{21}\left(4S_{12}S_{23}s_{13}+K_{12}(2c_{13}^{2}-(3-K_{13})K_{23}\right)\right],\\
H_{0}^{2} & =H_{0}^{5}=H_{0}^{7}=H_{0}^{0}=0\, .
\end{aligned}
\label{eq:16}
\end{equation}
}
.

The electron interaction term is 
\begin{equation}
H_{e}=\frac{\sqrt{2}A}{8}\left(3\lambda_{3}+\sqrt{3}\lambda_{8}\right).
\end{equation}
One defines the self-interaction strength as
\begin{equation}
B_{\gamma}(\vec{p},\vec{r},t)=\frac{\sqrt{2}G_{F}}{\hbar}\int d^{3}q(1-\cos\theta^{pq})f_{\gamma}(\vec{q},\vec{r},t)\, .
\end{equation}
Denoting anti-neutrino quantities with a bar, the self-interaction
Hamiltonians are written as
\begin{equation}
\begin{aligned}H_{\nu\nu} & =\sum_{\gamma=0}^{3}B_{\gamma}\sigma^{\gamma}\\
H_{\nu\bar{\nu}}^{*} & =\sum_{\gamma=0}^{3}\bar{B}_{\gamma}^{*}\sigma^{\gamma}
\end{aligned}
,
\end{equation}
and 
\begin{equation}
\begin{aligned}H_{\nu\nu} & =\sum_{\gamma=0}^{8}B_{\gamma}\lambda^{\gamma}\\
H_{\nu\bar{\nu}}^{*} & =\sum_{\gamma=0}^{8}\bar{B}_{\gamma}^{*}\lambda^{\gamma}
\end{aligned}
,
\end{equation}
for two and three flavors, respectively.

\subsection{The Matrix Equations\label{sub:The-Matrix-Equations}}

The transport equation in matrix form thus reads for two flavors,
summing from 0 to 3:
\begin{equation}
\begin{alignedat}{1}\frac{\partial\mathcal{F}}{\partial t}+\vec{v}\cdot\frac{\partial\mathcal{F}}{\partial\vec{r}}+\dot{\vec{p}}\cdot\frac{\partial\mathcal{F}}{\partial\vec{p}} & =2\epsilon_{\alpha\beta\gamma}(H_{0}^{\alpha}+H_{e}^{\alpha}+B^{\alpha}-\bar{B}^{\alpha*})f^{\beta}\sigma^{\gamma}+C,\\
\frac{\partial\bar{\mathcal{F}}}{\partial t}+\vec{v}\cdot\frac{\partial\bar{\mathcal{F}}}{\partial\vec{r}}+\dot{\vec{p}}\cdot\frac{\partial\bar{\mathcal{F}}}{\partial\vec{p}} & =2\epsilon_{\alpha\beta\gamma}(H_{0}^{\alpha}-H_{e}^{\alpha}-B^{\alpha*}+\bar{B}^{\alpha})\bar{f}^{\beta}\sigma^{\gamma}+\bar{C},
\end{alignedat}
\label{eq:22}
\end{equation}
where $\epsilon_{\alpha\beta\gamma}$ is the anti-symmetric tensor.
Again throughout this paper, anti-neutrino quantities are denoted
with a bar. Note that the anti-symmetric tensor induces a cross product:
\[
\begin{aligned}\vec{z}=\vec{x}\times\vec{y} & \ \ \ \Longleftrightarrow & z_{i}=\epsilon_{ijk}x^{j}y^{k}\end{aligned}
.
\]
Thus, Eq. \eqref{eq:22} contains the ``flavor pendulum" equation of motion,
and this cross-product form motivated the language used to describe collective neutrino
oscillations (see, for example, \citep{2010ARNPS..60..569D,2009NuPhS.188..101S,2013PhRvD..88d5031R} and references therein). 
(For more on the pendulum analogy, we refer the reader to Appendix \ref{sec:Full-SU(N)}.)

Similarly for three flavors, summing from 0 to 8
\begin{equation}
\begin{alignedat}{1}\frac{\partial\mathcal{F}}{\partial t}+\vec{v}\cdot\frac{\partial\mathcal{F}}{\partial\vec{r}}+\dot{\vec{p}}\cdot\frac{\partial\mathcal{F}}{\partial\vec{p}} & =2c_{\alpha\beta\gamma}(H_{0}^{\alpha}+H_{e}^{\alpha}+B^{\alpha}-\bar{B}^{\alpha*})f^{\beta}\lambda^{\gamma}+C,\\
\frac{\partial\bar{\mathcal{F}}}{\partial t}+\vec{v}\cdot\frac{\partial\bar{\mathcal{F}}}{\partial\vec{r}}+\dot{\vec{p}}\cdot\frac{\partial\bar{\mathcal{F}}}{\partial\vec{p}} & =2c_{\alpha\beta\gamma}(H_{0}^{\alpha}-H_{e}^{\alpha}-B^{\alpha*}+\bar{B}^{\alpha})\bar{f}^{\beta}\lambda^{\gamma}+\bar{C},
\end{alignedat}
\label{eq:23}
\end{equation}
where $c_{\alpha\beta\gamma}$ are the SU(3) structure constants given
in Appendix \ref{sec:Expansion}. The SU(3) structure can be similarly 
understood as a generalized cross product.  Eq. \eqref{eq:23} can be expanded
in the natural basis with Eqs. \eqref{eq:6}, \eqref{eq:8}, and \eqref{eq:43}.
For two flavors, Eq. \eqref{eq:22} can be similarly expanded.

\subsection{The Transport Basis\label{sub:The-Transport-Basis}}

In transport algorithms, it is easiest to keep the diagonal terms
in the natural basis, or flavor basis, and put the off-diagonal terms
in the SU(N) basis. We call this basis the ``transport basis.''
%%\footnote{The equations in full SU(N) basis are given in Appendix \ref{sec:Full-SU(N)}. %
%%}
$C$ and $\bar{C}$ are diagonal and, hence, unchanged. $\mathcal{F}$
is given in SU(N) expansion in Eqs.\eqref{eq:11} and \eqref{eq:12}. Notice
that $\sigma_{0,3}$ and $\lambda_{0,3,8}$ are diagonal, and, therefore,
we take linear combinations of them to restore the diagonal terms
to the flavor basis. 
For two flavors,
\begin{equation}
f_{0}=\frac{f_{\nu_{e}}+f_{\nu_{\mu}}}{2},\ f_{3}=\frac{f_{\nu_{e}}-f_{\nu_{\mu}}}{2},\label{eq:25}
\end{equation}
or inversely
\begin{equation}
f_{\nu_{e}}=f_{0}+f_{3},\ f_{\nu_{\mu}}=f_{0}-f_{3}\, .\label{eq:26}
\end{equation}
For three flavors, the relations are 
\begin{equation}
\begin{aligned}f_{0} & =\frac{1}{3}(f_{\nu_{e}}+f_{\nu_{\mu}}+f_{\nu_{\tau}}),\\
f_{3} & =\frac{1}{2}(f_{\nu_{e}}-f_{\nu_{\mu}}),\\
f_{8} & =\frac{1}{6\sqrt{3}}(f_{\nu_{e}}+f_{\nu_{\mu}}-2f_{\nu_{\tau}}),
\end{aligned}
\label{eq:27}
\end{equation}
or 
\begin{equation}
\begin{aligned}f_{\nu_{e}} & =f_{0}+f_{3}+\sqrt{3}f_{8},\\
f_{\nu_{\mu}} & =f_{0}-f_{3}+\sqrt{3}f_{8},\\
f_{\nu_{\tau}} & =f_{0}-2\sqrt{3}f_{8}\, .
\end{aligned}
\label{eq:28}
\end{equation}
The same form also applies to any other diagonal quantities
such as self-interaction strength $B$, or $C$ %
\footnote{The transformation for $C$ is used in the full SU(N) expansion; see
Appendix \ref{sec:Full-SU(N)}.%
} 
.

The two-flavor transport equations can, thus, be written as, noting
$\epsilon_{\alpha\beta0}=0$:
\begin{equation}
\begin{alignedat}{1}\frac{\partial f_{\nu_{e}}}{\partial t}+\vec{v}\cdot\frac{\partial f_{\nu_{e}}}{\partial\vec{r}}+\dot{\vec{p}}\cdot\frac{\partial f_{\nu_{e}}}{\partial\vec{p}} & =2\epsilon_{\alpha\beta3}(H_{0}^{\alpha}+H_{e}^{\alpha}+B^{\alpha}-\bar{B}^{\alpha*})f^{\beta}+C_{\nu_{e}},\\
\frac{\partial f_{\nu_{\mu}}}{\partial t}+\vec{v}\cdot\frac{\partial f_{\nu_{\mu}}}{\partial\vec{r}}+\dot{\vec{p}}\cdot\frac{\partial f_{\nu_{\mu}}}{\partial\vec{p}} & =-2\epsilon_{\alpha\beta3}(H_{0}^{\alpha}+H_{e}^{\alpha}+B^{\alpha}-\bar{B}^{\alpha*})f^{\beta}+C_{\nu_{\mu}},\\
\frac{\partial f_{\gamma}}{\partial t}+\vec{v}\cdot\frac{\partial f_{\gamma}}{\partial\vec{r}}+\dot{\vec{p}}\cdot\frac{\partial f_{\gamma}}{\partial\vec{p}} & =2\epsilon_{\alpha\beta\gamma}(H_{0}^{\alpha}+H_{e}^{\alpha}+B^{\alpha}-\bar{B}^{\alpha*})f^{\beta},\ \ \ \text{for }\gamma=1,2\, ,
\end{alignedat}
\end{equation}
and for anti-neutrinos

\begin{equation}
\begin{alignedat}{1}\frac{\partial\bar{f}_{\nu_{e}}}{\partial t}+\vec{v}\cdot\frac{\partial\bar{f}_{\nu_{e}}}{\partial\vec{r}}+\dot{\vec{p}}\cdot\frac{\partial\bar{f}_{\nu_{e}}}{\partial\vec{p}} & =2\epsilon_{\alpha\beta3}(H_{0}^{\alpha}-H_{e}^{\alpha}-B^{\alpha*}+\bar{B}^{\alpha})\bar{f}^{\beta}+\bar{C}_{\nu_{e}},\\
\frac{\partial\bar{f}_{\nu_{\mu}}}{\partial t}+\vec{v}\cdot\frac{\partial\bar{f}_{\nu_{\mu}}}{\partial\vec{r}}+\dot{\vec{p}}\cdot\frac{\partial\bar{f}_{\nu_{\mu}}}{\partial\vec{p}} & =2\epsilon_{\alpha\beta3}(H_{0}^{\alpha}-H_{e}^{\alpha}-B^{\alpha*}+\bar{B}^{\alpha})\bar{f}^{\beta}+\bar{C}_{\nu_{\mu}},\\
\frac{\partial\bar{f}_{\gamma}}{\partial t}+\vec{v}\cdot\frac{\partial\bar{f}_{\gamma}}{\partial\vec{r}}+\dot{\vec{p}}\cdot\frac{\partial\bar{f}_{\gamma}}{\partial\vec{p}} & =2\epsilon_{\alpha\beta\gamma}(H_{0}^{\alpha}-H_{e}^{\alpha}-B^{\alpha*}+\bar{B}^{\alpha})\bar{f}^{\beta},\ \ \ \text{for }\gamma=1,2\, .
\end{alignedat}
\end{equation}
The two-flavor equations expanded in component form are given in Appendix
\ref{sec:Expanded-two-flavor}. 
The three-flavor equations in the transport basis are, noting that $c_{\alpha\beta0}=0$
or, equivalently, that $f_{0}=(f_{\nu_{e}}+f_{\nu_{\mu}}+f_{\nu_{\tau}})/3$,
is not affected by oscillation:

\begin{equation}
\begin{aligned}\frac{\partial f_{\nu_{e}}}{\partial t}+\vec{v}\cdot\frac{\partial f_{\nu_{e}}}{\partial\vec{r}}+\dot{\vec{p}}\cdot\frac{\partial f_{\nu_{e}}}{\partial\vec{p}} & =2(c_{\alpha\beta3}+\sqrt{3}c_{\alpha\beta8})(H_{0}^{\alpha}+H_{e}^{\alpha}+B^{\alpha}-\bar{B}^{\alpha*})f^{\beta}+C_{\nu_{e}},\\
\frac{\partial f_{\nu_{\mu}}}{\partial t}+\vec{v}\cdot\frac{\partial f_{\nu_{\mu}}}{\partial\vec{r}}+\dot{\vec{p}}\cdot\frac{\partial f_{\nu_{\mu}}}{\partial\vec{p}} & =2(-c_{\alpha\beta3}+\sqrt{3}c_{\alpha\beta8})(H_{0}^{\alpha}+H_{e}^{\alpha}+B^{\alpha}-\bar{B}^{\alpha*})f^{\beta}+C_{\nu_{\mu}},\\
\frac{\partial f_{\nu_{\tau}}}{\partial t}+\vec{v}\cdot\frac{\partial f_{\nu_{\tau}}}{\partial\vec{r}}+\dot{\vec{p}}\cdot\frac{\partial f_{\nu_{\tau}}}{\partial\vec{p}} & =2(-2\sqrt{3}c_{\alpha\beta8})(H_{0}^{\alpha}+H_{e}^{\alpha}+B^{\alpha}-\bar{B}^{\alpha*})f^{\beta}+C_{\nu_{\tau}},\\
\frac{\partial f_{\gamma}}{\partial t}+\vec{v}\cdot\frac{\partial f_{\gamma}}{\partial\vec{r}}+\dot{\vec{p}}\cdot\frac{\partial f_{\gamma}}{\partial\vec{p}} & =2c_{\alpha\beta\gamma}(H_{0}^{\alpha}+H_{e}^{\alpha}+B^{\alpha}-\bar{B}^{\alpha*})f^{\beta},\ \ \ \text{for }\gamma=1,2,4,5,6,7\, .
\end{aligned}
\label{eq:29}
\end{equation}
and 
\begin{equation}
\begin{aligned}\frac{\partial\bar{f}_{\nu_{e}}}{\partial t}+\vec{v}\cdot\frac{\partial\bar{f}_{\nu_{e}}}{\partial\vec{r}}+\dot{\vec{p}}\cdot\frac{\partial\bar{f}_{\nu_{e}}}{\partial\vec{p}} & =2(c_{\alpha\beta3}+\sqrt{3}c_{\alpha\beta8})(H_{0}^{\alpha}-H_{e}^{\alpha}-B^{\alpha*}+\bar{B}^{\alpha})\bar{f}^{\beta}+\bar{C}_{\nu_{e}},\\
\frac{\partial\bar{f}_{\nu_{\mu}}}{\partial t}+\vec{v}\cdot\frac{\partial\bar{f}_{\nu_{\mu}}}{\partial\vec{r}}+\dot{\vec{p}}\cdot\frac{\partial\bar{f}_{\nu_{\mu}}}{\partial\vec{p}} & =2(-c_{\alpha\beta3}+\sqrt{3}c_{\alpha\beta8})(H_{0}^{\alpha}-H_{e}^{\alpha}-B^{\alpha*}+\bar{B}^{\alpha})\bar{f}^{\beta}+\bar{C}_{\nu_{\mu}},\\
\frac{\partial\bar{f}_{\nu_{\tau}}}{\partial t}+\vec{v}\cdot\frac{\partial\bar{f}_{\nu_{\tau}}}{\partial\vec{r}}+\dot{\vec{p}}\cdot\frac{\partial\bar{f}_{\nu_{\tau}}}{\partial\vec{p}} & =2(-2\sqrt{3}c_{\alpha\beta8})(H_{0}^{\alpha}-H_{e}^{\alpha}-B^{\alpha*}+\bar{B}^{\alpha})\bar{f}^{\beta}+\bar{C}_{\nu_{\tau}},\\
\frac{\partial\bar{f}_{\gamma}}{\partial t}+\vec{v}\cdot\frac{\partial\bar{f}_{\gamma}}{\partial\vec{r}}+\dot{\vec{p}}\cdot\frac{\partial\bar{f}_{\gamma}}{\partial\vec{p}} & =2c_{\alpha\beta\gamma}(H_{0}^{\alpha}-H_{e}^{\alpha}-B^{\alpha*}+\bar{B}^{\alpha})\bar{f}^{\beta},\ \ \ \text{for }\gamma=1,2,4,5,6,7\, .
\end{aligned}
\label{eq:30}
\end{equation}
Notice the last line in each of Eqs. \eqref{eq:29} and \eqref{eq:30} are the equations for 
the off-diagonal terms. (For three flavors, we here do not provide the
long form of the equation in components.)

\section{Moment Equations \label{sec:Moment-Equations}}

We let the specific intensity and angular moments have 
their usual definitions \citep{Mihalas1999,Castor2004}, summarized in Table II.
The nth-moment equation is obtained by integrating the transport equation
over $\int d\Omega_{p}\ \hat{p}^{n}$, where $\hat{p}$ is a unit
vector along the direction of momentum and $\Omega_{p}$ denotes solid angle. 
The scattering, absorption, and emission source terms can
be written in simple forms and are discussed in Appendix \ref{sec:Scattering-and-absorption}.
Such a set of equations contains the $n+1$th moment. Therefore, more information relating
the moments is needed, and this is called the closure problem \citep{Mihalas1999,Castor2004}. Typically a closure
among the 0th, 1st and 2nd moments is used, and we leave a discussion of
possible closures to future work.  
\footnote{It is straightforward to show that the moment formalism naturally
contains a geometric closure, consistent with the result of \citep{PhysRevD.78.033014},
when the single-angle approximation is assumed \citep{2011PhRvD..84c3013M}.  %  2013arXiv1302.1159C}. %
}

For two flavors, setting $\beta=\frac{\sqrt{2}G_{F}}{\hbar}$, we write the 0th and 1st moment equations in the transport
basis:
\begin{equation}
\begin{alignedat}{1}\frac{\partial E^{\nu_{e}}}{\partial t}+\partial^{j}F_{j}^{\nu_{e}}+\dot{p}^{j}\frac{\partial_{p}}{c}F_{j}^{\nu_{e}} & =2\epsilon_{\alpha\beta3}\left[(H_{0}^{\alpha}+H_{e}^{\alpha})E^{\beta}+\frac{\beta}{c}\int\frac{dq}{q}\left\{ c^{2}\breve{E}^{\alpha}(q)E^{\beta}-\breve{F_{i}^{\alpha}}(q)\cdot F_{i}^{\beta}\right\} \right]+C_{\nu_{e}}^{(0)},\\
\frac{\partial E^{\nu_{\mu}}}{\partial t}+\partial^{j}F_{j}^{\nu_{\mu}}+\dot{p}^{j}\frac{\partial_{p}}{c}F_{j}^{\nu_{\mu}} & =-2\epsilon_{\alpha\beta3}\left[(H_{0}^{\alpha}+H_{e}^{\alpha})E^{\beta}+\frac{\beta}{c}\int\frac{dq}{q}\left\{ c^{2}\breve{E}^{\alpha}(q)E^{\beta}-\breve{F_{i}^{\alpha}}(q)\cdot F_{i}^{\beta}\right\} \right]+C_{\nu_{\mu}}^{(0)},\\
\frac{\partial E^{\gamma}}{\partial t}+\partial^{j}F_{j}^{\gamma}+\dot{p}^{j}\frac{\partial_{p}}{c}F_{j}^{\gamma} & =2\epsilon_{\alpha\beta\gamma}\left[(H_{0}^{\alpha}+H_{e}^{\alpha})E^{\beta}+\frac{\beta}{c}\int\frac{dq}{q}\left\{ c^{2}\breve{E}^{\alpha}(q)E^{\beta}-\breve{F_{i}^{\alpha}}(q)\cdot F_{i}^{\beta}\right\} \right],\ \ \text{for }\gamma=1,2\, ,
\end{alignedat}
\end{equation}
and 
\begin{equation}
\begin{alignedat}{1}\frac{\partial F_{i}^{\nu_{e}}}{\partial t}+\partial^{j}P_{ij}^{\nu_{e}}+\dot{p}^{j}\frac{\partial_{p}}{c}P_{ij}^{\nu_{e}} & =2c_{\alpha\beta3}\left[(H_{0}^{\alpha}+H_{e}^{\alpha})F_{i}^{\beta}+\beta\int\frac{dq}{q}\left\{ \breve{E}^{\alpha}(q)F_{i}^{\beta}-\breve{F}_{j}^{\alpha}(q)\cdot P_{ij}^{\beta}\right\} \right]+C_{\nu_{e}}^{(1)},\\
\frac{\partial F_{i}^{\nu_{\mu}}}{\partial t}+\partial^{j}P_{ij}^{\nu_{\mu}}+\dot{p}^{j}\frac{\partial_{p}}{c}P_{ij}^{\nu_{\mu}} & =-2\epsilon_{\alpha\beta3}\left[(H_{0}^{\alpha}+H_{e}^{\alpha})F_{i}^{\beta}+\beta\int\frac{dq}{q}\left\{ \breve{E}^{\alpha}(q)F_{i}^{\beta}-\breve{F}_{j}^{\alpha}(q)\cdot P_{ij}^{\beta}\right\} \right]+C_{\nu_{\mu}}^{(1)},\\
\frac{\partial F_{i}^{\gamma}}{\partial t}+\partial^{j}P_{ij}^{\gamma}+\dot{p}^{j}\frac{\partial_{p}}{c}P_{ij}^{\gamma} & =2\epsilon_{\alpha\beta\gamma}\left[(H_{0}^{\alpha}+H_{e}^{\alpha})F_{i}^{\beta}+\beta\int\frac{dq}{q}\left\{ \breve{E}^{\alpha}(q)F_{i}^{\beta}-\breve{F}_{j}^{\alpha}(q)\cdot P_{ij}^{\beta}\right\} \right],\ \text{for }\gamma=1,2\, ,
\end{alignedat}
\end{equation}
where $C_{\nu_{e,\mu}}^{(0,1)}$ are the moments of the scattering
and absorption terms, discussed in Appendix \ref{sec:Scattering-and-absorption}.
The $\frac{\partial_{p}}{c}$ on the left hand side is differentiation
with respect to energy and again is related to gravitational redshift.
The breve ($\breve{}$) is a short hand for
\begin{equation}
\begin{aligned}\breve{E}_{\alpha} & =E_{\alpha}(q)-\bar{E}_{\alpha}^{*}(q),\\
\breve{\vec{F}}_{\alpha} & =\vec{F}_{\alpha}(q)-\bar{\vec{F}}_{\alpha}^{*}(q)\, .
\end{aligned}
\end{equation}
$E$ and $F$ are the generalized neutrino energy density and flux spectra: the 0th and 1st 
angular moments of the compenents of the distribution function/density matrix. 
For anti-neutrinos, the equations are obtained by the substitutions:

\begin{equation}
\begin{aligned}H_{e} & \rightarrow-H_{e},\\
E_{\alpha}(q)-\bar{E}_{\alpha}^{*}(q), & \rightarrow-E_{\alpha}^{*}(q)+\bar{E}_{\alpha}(q),\\
\vec{F}_{\alpha}(q)-\bar{\vec{F}}_{\alpha}^{*}(q) & \rightarrow-\vec{F}_{\alpha}^{*}(q)+\bar{\vec{F}}_{\alpha}(q).
\end{aligned}
\label{eq:33}
\end{equation}
For the full expansion in component form, we refer the reader
to Appendix \ref{sec:Expanded-two-flavor}. 

For three flavors, the first two transport moment equations in the transport
basis are:

\begin{equation}
\begin{alignedat}{1}\frac{\partial E^{\nu_{e}}}{\partial t}+\partial^{j}F_{j}^{\nu_{e}}+\dot{p}^{j}\frac{\partial_{p}}{c}F_{j}^{\nu_{e}} & =2(c_{\alpha\beta3}+\sqrt{3}c_{\alpha\beta8})\left[(H_{0}^{\alpha}+H_{e}^{\alpha})E^{\beta}+\frac{\beta}{c}\int\frac{dq}{q}\left\{ c^{2}\breve{E}^{\alpha}(q)E^{\beta}-\breve{F_{i}^{\alpha}}(q)\cdot F_{i}^{\beta}\right\} \right]+C_{\nu_{e}}^{(0)},\\
\frac{\partial E^{\nu_{\mu}}}{\partial t}+\partial^{j}F_{j}^{\nu_{\mu}}+\dot{p}^{j}\frac{\partial_{p}}{c}F_{j}^{\nu_{\mu}} & =2(-c_{\alpha\beta3}+\sqrt{3}c_{\alpha\beta8})\left[(H_{0}^{\alpha}+H_{e}^{\alpha})E^{\beta}+\frac{\beta}{c}\int\frac{dq}{q}\left\{ c^{2}\breve{E}^{\alpha}(q)E^{\beta}-\breve{F_{i}^{\alpha}}(q)\cdot F_{i}^{\beta}\right\} \right]+C_{\nu_{\mu}}^{(0)},\\
\frac{\partial E^{\nu_{\tau}}}{\partial t}+\partial^{j}F_{j}^{\nu_{\tau}}+\dot{p}^{j}\frac{\partial_{p}}{c}F_{j}^{\nu_{\tau}} & =2(-2\sqrt{3}c_{\alpha\beta8})\left[(H_{0}^{\alpha}+H_{e}^{\alpha})E^{\beta}+\frac{\beta}{c}\int\frac{dq}{q}\left\{ c^{2}\breve{E}^{\alpha}(q)E^{\beta}-\breve{F_{i}^{\alpha}}(q)\cdot F_{i}^{\beta}\right\} \right]+C_{\nu_{\tau}}^{(0)},\\
\frac{\partial E^{\gamma}}{\partial t}+\partial^{j}F_{j}^{\gamma}+\dot{p}^{j}\frac{\partial_{p}}{c}F_{j}^{\gamma} & =2c_{\alpha\beta\gamma}\left[(H_{0}^{\alpha}+H_{e}^{\alpha})E^{\beta}+\frac{\beta}{c}\int\frac{dq}{q}\left\{ c^{2}\breve{E}^{\alpha}(q)E^{\beta}-\breve{F_{i}^{\alpha}}(q)\cdot F_{i}^{\beta}\right\} \right],\ \ \text{for }\gamma=1,2,4,5,6,7,
\end{alignedat}
\end{equation}
and 
\begin{equation}
\begin{alignedat}{1}\frac{\partial F_{i}^{\nu_{e}}}{\partial t}+\partial^{j}P_{ij}^{\nu_{e}}+\dot{p}^{j}\frac{\partial_{p}}{c}P_{ij}^{\nu_{e}} & =2(c_{\alpha\beta3}+\sqrt{3}c_{\alpha\beta8})\left[(H_{0}^{\alpha}+H_{e}^{\alpha})F_{i}^{\beta}+\beta\int\frac{dq}{q}\left\{ \breve{E}^{\alpha}(q)F_{i}^{\beta}-\breve{F}_{j}^{\alpha}(q)\cdot P_{ij}^{\beta}\right\} \right]+C_{\nu_{e}}^{(1)},\\
\frac{\partial F_{i}^{\nu_{\mu}}}{\partial t}+\partial^{j}P_{ij}^{\nu_{\mu}}+\dot{p}^{j}\frac{\partial_{p}}{c}P_{ij}^{\nu_{\mu}} & =2(-c_{\alpha\beta3}+\sqrt{3}c_{\alpha\beta8})\left[(H_{0}^{\alpha}+H_{e}^{\alpha})F_{i}^{\beta}+\beta\int\frac{dq}{q}\left\{ \breve{E}^{\alpha}(q)F_{i}^{\beta}-\breve{F}_{j}^{\alpha}(q)\cdot P_{ij}^{\beta}\right\} \right]+C_{\nu_{\mu}}^{(1)},\\
\frac{\partial F_{i}^{\nu_{\tau}}}{\partial t}+\partial^{j}P_{ij}^{\nu_{\tau}}+\dot{p}^{j}\frac{\partial_{p}}{c}P_{ij}^{\nu_{\tau}} & =2(-2\sqrt{3}c_{\alpha\beta8})\left[(H_{0}^{\alpha}+H_{e}^{\alpha})F_{i}^{\beta}+\beta\int\frac{dq}{q}\left\{ \breve{E}^{\alpha}(q)F_{i}^{\beta}-\breve{F}_{j}^{\alpha}(q)\cdot P_{ij}^{\beta}\right\} \right]+C_{\nu_{\tau}}^{(1)},\\
\frac{\partial F_{i}^{\gamma}}{\partial t}+\partial^{j}P_{ij}^{\gamma}+\dot{p}^{j}\frac{\partial_{p}}{c}P_{ij}^{\gamma} & =2c_{\alpha\beta\gamma}\left[(H_{0}^{\alpha}+H_{e}^{\alpha})F_{i}^{\beta}+\beta\int\frac{dq}{q}\left\{ \breve{E}^{\alpha}(q)F_{i}^{\beta}-\breve{F}_{j}^{\alpha}(q)\cdot P_{ij}^{\beta}\right\} \right],\ \text{for }\gamma=1,2,4,5,6,7.
\end{alignedat}
\end{equation}

\section{Conclusions}
\label{conclusion}

In this paper, we have provided a suite of generalized Boltzmann equations, based on the density-matrix
formalism, that incorporates the physics of neutrino oscillations for two- and three-flavor oscillations, matter refraction, and self-refraction.
The resulting equations are straightforward extensions of the classical transport equations that by
their cross couplings and augmentation to include off-diagonal densities are of a form usefully similar 
to the classical, diagonal flavor-basis set of partial differential equations.  The expanded equation set nevertheless
contains the full physics of quantum oscillation phenomena, though it maintains the classical format
familiar to supernova astrophysicists who perform traditional neutrino transport simulations.  In this way, 
our formalism provides a bridge between the familiar approaches employed by supernova modelers and 
the formalisms employed by the pioneers in neutrino oscillation physics.
We also provide the corresponding angular-moment versions of this generalized equation set.
Our goal is to make it easier for astrophysicists to address oscillation phenomena in a language with which 
they are familiar.  At the same time, we hope that neutrino oscillation experts interested
in incorporating the effects of transport in a natural way may find our formalism of use.
While we have not included a discussion of sterile neutrinos or spin flips \citep{2013arXiv1309.2628V},
and have not explored the differences between Majorana and Dirac \citep{1992PhRvD..46..510P} neutrinos, nor 
possible neutrino-antineutrino oscillations \citep{2013arXiv1309.2628V}, we believe the equations derived 
are simple, clear, and practical renditions that will facilitate progress on oscillation phenomena 
in the context of core-collapse theory. 

\begin{acknowledgments}
A.B. acknowledges support by the Scientific Discovery through
Advanced Computing (SciDAC) program of the DOE, under grant number DE-FG02-08ER41544,
the NSF under the subaward no. ND201387 to the Joint Institute for Nuclear Astrophysics (JINA, NSF PHY-0822648),
and the NSF PetaApps program, under award OCI-0905046 via a subaward
no. 44592 from Louisiana State University to Princeton University.
\end{acknowledgments}

\appendix
\[
\]

\section{SU(2) and SU(3) \label{sec:Expansion}}

The Dirac matrices 
\begin{equation}
\begin{aligned}\sigma^{0}=\left(\begin{array}{cc}
1 & 0\\
0 & 1
\end{array}\right)\ ,\ \sigma^{1}=\left(\begin{array}{cc}
0 & 1\\
1 & 0
\end{array}\right)\ ,\  & \sigma^{2}=\left(\begin{array}{cc}
0 & -i\\
i & 0
\end{array}\right)\ ,\  & \sigma^{3}=\left(\begin{array}{cc}
1 & 0\\
0 & -1
\end{array}\right)\end{aligned}
\end{equation}
satisfy the SU(2) commutation relations
\begin{equation}
\left[\frac{\sigma_{\alpha}}{2},\frac{\sigma_{\beta}}{2}\right]=\epsilon_{\alpha\beta\gamma}\frac{\sigma^{\gamma}}{2}\, ,
\end{equation}
where the structure constants are represented by the anti-symmetric
tensor $\epsilon_{\alpha\beta\gamma}$, which in particular vanishes
when any of the indices are zero. 

The Gell-Mann matrices
\begin{equation}
\begin{alignedat}{2}\lambda_{0}=\left[\begin{array}{ccc}
1 & 0 & 0\\
0 & 1 & 0\\
0 & 0 & 1
\end{array}\right], & \lambda_{1}=\left[\begin{array}{ccc}
0 & 1 & 0\\
1 & 0 & 0\\
0 & 0 & 0
\end{array}\right], & \lambda_{2}=\left[\begin{array}{ccc}
0 & -i & 0\\
i & 0 & 0\\
0 & 0 & 0
\end{array}\right],\\
\lambda_{3}=\left[\begin{array}{ccc}
1 & 0 & 0\\
0 & -1 & 0\\
0 & 0 & 0
\end{array}\right], & \lambda_{4}=\left[\begin{array}{ccc}
0 & 0 & 1\\
0 & 0 & 0\\
1 & 0 & 0
\end{array}\right], & \lambda_{5}=\left[\begin{array}{ccc}
0 & 0 & -i\\
0 & 0 & 0\\
i & 0 & 0
\end{array}\right],\\
\lambda_{6}=\left[\begin{array}{ccc}
0 & 0 & 0\\
0 & 0 & 1\\
0 & 1 & 0
\end{array}\right], & \lambda_{7}=\left[\begin{array}{ccc}
0 & 0 & 0\\
0 & 0 & -i\\
0 & i & 0
\end{array}\right], & \lambda_{8}=\frac{1}{\sqrt{3}}\left[\begin{array}{ccc}
1 & 0 & 0\\
0 & 1 & 0\\
0 & 0 & -2
\end{array}\right],
\end{alignedat}
\label{eq:43}
\end{equation}
satisfy the SU(3) commutation 
\begin{equation}
\left[\frac{\lambda_{\alpha}}{2},\frac{\lambda_{\beta}}{2}\right]=ic_{\alpha\beta\gamma}\frac{\lambda_{\gamma}}{2}\, .
\end{equation}
The structure constants $c_{\alpha\beta\gamma}$ are anti-symmetric
with respect to exchange of pair indices, and in particular $c_{\alpha\beta0}=0$.
The non-vanishing components can be specified via
\begin{equation}
c_{123}=2;\ c_{147},c_{165},c_{246},c_{257},c_{345},c_{376}=1;\ c_{678},c_{458}=\sqrt{3}.
\end{equation}

\section{Boltzmann Equations in the Full SU(N) Basis\label{sec:Full-SU(N)}}

The flavor pendulum equation of motion widely discussed in the literature
is equivalent to expanding the matrix equation in full SU(N) basis.
To write the diagonal terms in the SU(N) basis requires the transformation
Eqs. \eqref{eq:25}-\eqref{eq:28}, with $f$ replaced by the source
terms $C$:
\begin{equation}
C_{0}=\frac{C_{\nu_{e}}+C_{\nu_{\mu}}}{2},\ C_{3}=\frac{C_{\nu_{e}}-C_{\nu_{\mu}}}{2},\label{eq:44-1}
\end{equation}
for two flavors and
\begin{equation}
\begin{aligned}C_{0} & =\frac{1}{3}(C_{\nu_{e}}+C_{\nu_{\mu}}+C_{\nu_{\tau}}),\\
C_{3} & =\frac{1}{2}(C_{\nu_{e}}-C_{\nu_{\mu}}),\\
C_{8} & =\frac{1}{6\sqrt{3}}(C_{\nu_{e}}+C_{\nu_{\mu}}-2C_{\nu_{\tau}}),
\end{aligned}
\label{eq:45-1}
\end{equation}
for three flavors. 

The resulting equations read
\begin{equation}
\begin{aligned}\frac{\partial f_{\gamma}}{\partial t}+\vec{v}\cdot\frac{\partial f_{\gamma}}{\partial\vec{r}}+\dot{\vec{p}}\cdot\frac{\partial f_{\gamma}}{\partial\vec{p}} & =2\chi_{\alpha\beta\gamma}(H_{0}^{\alpha}+H_{e}^{\alpha}+B^{\alpha}-\bar{B}^{\alpha*})f^{\beta}+C_{\gamma},\\
\frac{\partial\bar{f}_{\gamma}}{\partial t}+\vec{v}\cdot\frac{\partial\bar{f}_{\gamma}}{\partial\vec{r}}+\dot{\vec{p}}\cdot\frac{\partial\bar{f}_{\gamma}}{\partial\vec{p}} & =2\chi_{\alpha\beta\gamma}(H_{0}^{\alpha}-H_{e}^{\alpha}-B^{\alpha*}+\bar{B}^{\alpha})\bar{f}^{\beta}+\bar{C}_{\gamma},
\end{aligned}
\label{eq:446}
\end{equation}
where 
\begin{equation}
\begin{cases}
\chi_{\alpha\beta\gamma}=\epsilon_{\alpha\beta\gamma} & \text{for two flavors,}\\
\chi_{\alpha\beta\gamma}=c_{\alpha\beta\gamma} & \text{for three flavors.}
\end{cases}
\end{equation}
The non-zero components of $C_{\gamma}$ are given by Eq. \eqref{eq:44-1}
and Eq. \eqref{eq:45-1}, similarly for $\bar{C}_{\gamma}$. 
Writing $f_{\gamma}$, $H_{0}^{\alpha}$ and so on as vectors, recognizing
$\chi_{\alpha\beta\gamma}$ as a cross product operator, and denoting
$D_{t}=\frac{\partial}{\partial t}+\vec{v}\cdot\frac{\partial}{\partial\vec{r}}+\dot{\vec{p}}\cdot\frac{\partial}{\partial\vec{p}}$,
we write Eq. \eqref{eq:446} as
\begin{equation}
\begin{aligned}D_{t}\vec{f} & =2(\vec{H}_{0}+\vec{H}_{e}+\vec{B}-\vec{\bar{B^{*}}})\times\vec{f}+\vec{C},\\
D_{t}\vec{\bar{f}} & =2(\vec{H}_{0}-\vec{H}_{e}-\vec{B^{*}}+\vec{\bar{B}})\times\vec{f}+\vec{C}.
\end{aligned}
\end{equation}
This is the flavor pendulum equation of motion, taking its name from 
the analogy with the equation of motion of a classical spinning top in a 
gravitational field, with an additional term ($\vec{C}$) for classical scattering,
absorption, and emission.

\section{Classical Source Terms \label{sec:Scattering-and-absorption}}

The scattering, absorption, and emission terms can be written \citep{2005PhRvD..71i3004S}, dropping flavor
indices, as:
\begin{equation}
\frac{C}{c\gamma}=\kappa^{a}\left(\frac{\Gamma-I}{1-f^{eq}}\right)-\kappa^{s}I+\frac{\kappa^{s}}{4\pi}\int\Phi(\Omega_{p},\Omega_{q})I(\Omega_{q})d\Omega_{q}\, ,\label{eq:44}
\end{equation}
where $a$ and $s$ denote absorption and scattering, respectively.
$\gamma$ is given in Table II.
$f^{eq}$ is the equilibrium Fermi-Dirac distribution for the flavor
in question. $\Gamma$ is the blackbody specific intensity. The blocking
factor $1-f^{eq}$ corrects for the filled states in fermion statistics,
a mechanism called stimulated absorption, discussed in
\citep{2006NuPhA.777..356B}. The coefficients are related to the
relevant cross sections $\sigma_{i}$ via 
\begin{equation}
\kappa=\sum_{i}n_{i}\sigma_{i}\ .\label{eq:45}
\end{equation}
$\Phi$ represents scattering back into the beam and can be approximated
by 
\begin{equation}
\Phi_{i}(\Omega_{p},\Omega_{q})=1+\delta_{i}\cos\theta_{pq}\, ,\label{eq:46}
\end{equation}
where in both formulae $i$ is an index for the type of scattering
or absorption, i.e. the type of particles involved in the scattering
or absorption.

With Eq. \eqref{eq:44}, Eq. \eqref{eq:45}, and Eq. \eqref{eq:46}, we have:
\begin{equation}
\frac{C}{\gamma c}=-\kappa^{s}I+\kappa^{a}\left(\frac{\Gamma-I}{1-F^{eq}}\right)+\frac{\kappa^{s}}{4\pi}\left[cE(q)+\delta_{T}\hat{p}\cdot\vec{F}(q)\right],
\end{equation}
where $\delta_{T}=\frac{\sum_{i}n_{i}\sigma_{i}\delta_{i}}{\sum_{i}n_{i}\sigma_{i}}.$
But for conservative scattering $\Phi$ is only a phase function of angles, and thus $|p|=|q|$. 
The generalization for inelastic scattering is straightforward.
The zeroth moment is 
\begin{equation}
\begin{aligned}\frac{1}{c\gamma}C^{(0)} & =-\kappa^{s}cE(p)+\kappa^{a}\left(\frac{4\pi\Gamma-cE(p)}{1-f^{eq}}\right)+\frac{\kappa^{s}}{4\pi}\left[4\pi cE(p)\right]\\
 & =\frac{\kappa^{a}}{1-f^{eq}}\left(4\pi\Gamma-cE(p)\right)\\
 & \equiv\kappa_{a}^{*}\left(4\pi\Gamma-cE(p)\right)\, .
\end{aligned}
\end{equation}
Notice that these are the usual transfer moment equation terms with source
$4\pi\Gamma\kappa^{a}$ and sink $-\kappa^{a}cE$, modulated by the 
blocking term $1/(1-f^{eq})$. 
The 1st moment is

\begin{equation}
\begin{aligned}\frac{1}{c\gamma}C^{(1)i} & =-\kappa^{s}F^{i}-\kappa^{a}\left(\frac{F^{i}}{1-f^{eq}}\right)+\frac{\kappa^{s}}{4\pi}\left[\delta_{T}\frac{4\pi\delta_{ij}}{3}F^{j}\right]\\
 & =-(\kappa^{s}+\frac{\kappa^{a}}{1-f^{eq}}-\frac{1}{3}\kappa^{s}\delta_{T})F^{i}\\
 & \equiv-\kappa_{T}F^{i}\, .
\end{aligned}
\end{equation}

\section{Components of Two-Flavor Transport Equations\label{sec:Expanded-two-flavor}}

For components, we use the following notation for any neutrino variable $Q$:
\begin{equation}
\begin{aligned}\hat{Q}_{2} & =Q_{2}+\bar{Q}_{2}\\
\breve{Q}_{1,3} & =Q_{1,3}-\bar{Q}_{1,3}\, ,
\end{aligned}
\end{equation}
where again the bar denotes the anti-neutrino.

Thus: 

\begin{equation}
%\begin{cases}
\begin{aligned}
\frac{\partial f_{\nu_{e}}}{\partial t}+v\cdot\frac{\partial f_{\nu_{e}}}{\partial r}+\dot{p}\cdot\frac{\partial f_{\nu_{e}}}{\partial p}=f_{2}\omega\sin2\theta+2\left[f_{2}\breve{B_{1}}-f_{1}\hat{B_{2}}\right]+C_{\nu_{e}},\\
\frac{\partial f_{\nu_{\mu}}}{\partial t}+v\cdot\frac{\partial f_{\nu_{\mu}}}{\partial r}+\dot{p}\cdot\frac{\partial f_{\nu_{\mu}}}{\partial p}=-f_{2}\omega\sin2\theta-2\left[f_{2}\breve{B_{1}}-f_{1}\hat{B_{2}}\right]+C_{\nu_{\mu}},\\
\frac{\partial f_{r}}{\partial t}+v\cdot\frac{\partial f_{r}}{\partial r}+\dot{p}\cdot\frac{\partial f_{r}}{\partial p}=-f_{2}\left[A-\omega\cos2\theta\right]-2\left[f_{2}\breve{B_{3}}-f_{3}\hat{B_{2}}\right],\\
\frac{\partial f_{i}}{\partial t}+v\cdot\frac{\partial f_{i}}{\partial r}+\dot{p}\cdot\frac{\partial f_{i}}{\partial p}=-f_{3}\omega\sin2\theta+f_{1}\left[A-\omega\cos2\theta\right]-2\left[f_{3}\breve{B_{1}}-f_{1}\breve{B_{3}}\right],
%%\end{cases}\label{eq:022}
\end{aligned}\label{eq:022}
\end{equation}
and 
\begin{equation}
%\begin{cases}
\begin{aligned}
\frac{\partial\bar{f}_{\nu_{e}}}{\partial t}+v\cdot\frac{\partial\bar{f}_{\nu_{e}}}{\partial r}+\dot{p}\cdot\frac{\partial\bar{f}_{\nu_{e}}}{\partial p}=\bar{f_{2}}\omega\sin2\theta-2\left[\bar{f_{2}}\breve{B_{1}}+\bar{f_{1}}\hat{B_{2}}\right]+\bar{C}_{\nu_{e}},\\
\frac{\partial\bar{f}_{\nu_{\mu}}}{\partial t}+v\cdot\frac{\partial\bar{f}_{\nu_{\mu}}}{\partial r}+\dot{p}\cdot\frac{\partial\bar{f}_{\nu_{\mu}}}{\partial p}=-\bar{f_{2}}\omega\sin2\theta+2\left[\bar{f_{2}}\breve{B_{1}}+\bar{f_{1}}\hat{B_{2}}\right]+\bar{C}_{\nu_{\mu}},\\
\frac{\partial\bar{f_{1}}}{\partial t}+v\cdot\frac{\partial\bar{f_{1}}}{\partial r}+\dot{p}\cdot\frac{\partial\bar{f_{1}}}{\partial p}=\bar{f_{2}}\left[A+\omega\cos2\theta\right]+2\left[\bar{f_{2}}\breve{B}_{3}+\bar{f_{3}}\hat{B}_{2}\right],\\
\frac{\partial\bar{f_{2}}}{\partial t}+v\cdot\frac{\partial\bar{f_{2}}}{\partial r}+\dot{p}\cdot\frac{\partial\bar{f_{2}}}{\partial p}=-\bar{f_{3}}\omega\sin2\theta+\bar{f_{1}}\left[-A-\omega\cos2\theta\right]+2\left[\bar{f_{3}}\breve{B_{1}}-\bar{f_{1}}\breve{B_{3}}\right]\, .
%%\end{cases}\label{eq:023}
\end{aligned}\label{eq:023}
\end{equation}

The corresponding first two-moment equations for neutrinos are 

\begin{equation}
\begin{alignedat}{1}\partial_{t}E_{\nu_{e}} & +\nabla\cdot\vec{F}_{\nu_{e}}+\frac{1}{c}\dot{\vec{p}}\cdot(\partial_{p}\vec{F}_{\nu_{e}})-\kappa_{a}^{*}\left(4\pi\Gamma_{\nu_{e}}-cE_{\nu_{e}}\right)\\
 & =E_{2}\omega\sin2\theta+2\frac{\beta}{c^{2}}\int\frac{dq}{q}(c^{2}\breve{E}_{1}(q)E_{2}-\breve{\vec{F}}_{1}(q)\cdot\vec{F}_{2}-c^{2}\hat{E_{2}}(q)E_{1}+\hat{\vec{F_{2}}}(q)\cdot\vec{F}_{1}),\\
\partial_{t}E_{\nu_{\mu}} & +\nabla\cdot\vec{F}_{\nu_{\mu}}+\frac{1}{c}\dot{\vec{p}}\cdot(\partial_{p}\vec{F}_{\nu_{\mu}})-\kappa_{a}^{*}\left(4\pi\Gamma_{\nu_{\mu}}-cE_{\nu_{\mu}}\right)\\
 & =-E_{i}\omega\sin2\theta-2\frac{\beta}{c^{2}}\int\frac{dq}{q}(c^{2}\breve{E}_{1}(q)E_{2}-\breve{\vec{F}}_{1}(q)\cdot\vec{F}_{2}-c^{2}\hat{E_{2}}(q)E_{1}+\hat{\vec{F_{2}}}(q)\cdot\vec{F}_{1}),\\
\partial_{t}E_{1} & +\nabla\cdot\vec{F_{1}}+\frac{1}{c}\dot{\vec{p}}\cdot(\partial_{p}\vec{F_{1}})\\
 & =-E_{2}(A-\omega\cos2\theta)-2\frac{\beta}{c^{2}}\int\frac{dq}{q}(c^{2}\breve{E}_{3}(q)E_{2}-\breve{\vec{F}}_{3}(q)\cdot\vec{F}_{2}-c^{2}\hat{E_{2}}(q)E_{3}+\hat{\vec{F_{2}}}(q)\cdot\vec{F}_{3}),\\
\partial_{t}E_{2} & +\nabla\cdot\vec{F_{2}}+\frac{1}{c}\dot{\vec{p}}\cdot(\partial_{p}\vec{F_{2}})\\
 & =-E_{3}\omega\sin2\theta+E_{1}\left(A-\omega\cos2\theta\right)-2\frac{\beta}{c^{2}}\int\frac{dq}{q}(c^{2}\breve{E}_{1}(q)E_{3}-\breve{\vec{F}}_{1}(q)\cdot\vec{F}_{3}-c^{2}\breve{E_{3}}(q)E_{1}+\breve{\vec{F_{3}}}(q)\cdot\vec{F}_{1})\, .
\end{alignedat}
\end{equation}
and
\begin{equation}
\begin{alignedat}{1}\partial_{t}F_{\nu_{e}}^{i} & +c^{2}\partial_{j}P_{\nu_{e}}^{ij}+\dot{p}^{j}\partial_{p}P_{\nu_{e}}^{ij}+c\kappa_{T}F_{\nu_{e}}^{i}\\
 & =F_{2}^{i}\omega\sin2\theta+2\beta\int\frac{dq}{q}(\breve{E}_{1}(q)F_{2}^{i}-\breve{F{}_{1}^{j}}(q)P_{2}^{ij}-\hat{E_{2}}(q)F_{1}^{i}+\hat{F_{2}^{j}}(q)P_{1}^{ij}),\\
\partial_{t}F_{\nu_{\mu}}^{i} & +c^{2}\partial_{j}P_{\nu_{\mu}}^{ij}+\dot{p}^{j}\partial_{p}P_{\nu_{\mu}}^{ij}+c\kappa_{T}F_{\nu_{\mu}}^{i}\\
 & =-F_{2}^{i}\omega\sin2\theta-2\beta\int\frac{dq}{q}(\breve{E}_{1}(q)F_{2}^{i}-\breve{F{}_{1}^{j}}(q)P_{2}^{ij}-\hat{E_{2}}(q)F_{1}^{i}+\hat{F_{2}^{j}}(q)P_{1}^{ij}),\\
\partial_{t}F_{1}^{i} & +c^{2}\partial_{j}P_{1}^{ij}+\dot{p}^{j}\partial_{p}P_{1}^{ij}\\
 & =-F_{2}^{i}(A-\omega\cos2\theta)-2\beta\int\frac{dq}{q}(\breve{E}_{3}(q)F_{2}^{i}-\breve{F{}_{3}^{j}}(q)P_{2}^{ij}-\hat{E_{2}}(q)F_{3}^{i}+\hat{F_{2}^{j}}(q)P_{3}^{ij}),\\
\partial_{t}F_{2}^{i} & +c^{2}\partial_{j}P_{2}^{ij}+\dot{p}^{j}\partial_{p}P_{2}^{ij}\\
 & =-F_{3}^{i}\omega\sin2\theta+cF_{2}^{i}\left(A-\omega\cos2\theta\right)-2\beta\int\frac{dq}{q}(\breve{E}_{1}(q)F_{3}^{i}-\breve{F{}_{1}^{j}}(q)P_{3}^{ij}-\breve{E_{3}}(q)F_{1}^{i}+\breve{F_{3}^{j}}(q)P_{1}^{ij})\, .
\end{alignedat}
\end{equation}

\bibliographystyle{apsrev}
\bibliography{try,books}

\newpage

\begin{center}
\begin{table}  %  [H]
\caption{\label{tab:Neutrino-Mixing-Parameters}Neutrino Mixing Parameters}
\begin{tabular}{ccc}
\toprule
$\Delta m^{2}$ & NH & IH  \tabularnewline
%\midrule
%\midrule
$\Delta m_{21}^{2}$ & $7.5\times10^{-5}eV^{2}$ & $7.5\times10^{-5}eV^{2}$\tabularnewline
$\Delta m_{32}^{2}$ & $2.39\times10^{-3}eV^{2}$ & $-2.42\times10^{-3}eV^{2}$\tabularnewline
$\Delta m_{31}^{2}$ & $2.47\times10^{-3}eV^{2}$ & $-2.34\times10^{-3}eV^{2}$\tabularnewline
%\bottomrule
\end{tabular}
\end{table}
%\end{center}

%\begin{center}
\begin{table}  %  [H]
\begin{tabular}{cc}
\toprule
Mixing angles & rad\tabularnewline
%\midrule
%\midrule
$\theta_{12}$ & 0.58\tabularnewline
$\theta_{23}$ & 0.67\tabularnewline
$\theta_{13}$ & 0.15\tabularnewline
%\bottomrule
\end{tabular}
\end{table}
\end{center}

%\begin{center}
\begin{table}  %  [H]
\caption{Definitions of crucial quantities}
\begin{tabular}{lll}
\toprule
Quantity  & Definition \footnote{$\varepsilon$ is the particle energy.} & Dimension\tabularnewline
%\midrule
%\midrule
Wigner Phase Space density & $f(r,p,t)=\int d^{3}R\ e^{-ipR}\psi^{\dagger}(r-\frac{R}{2},t)\psi(r+\frac{R}{2},t)$ & 1\tabularnewline
Specific Intensity  & $I_{\varepsilon}=\frac{\varepsilon^{3}f}{c^{2}(2\pi\hbar)^{3}}=cp^{3}\frac{f}{(2\pi\hbar)^{3}}\equiv\gamma^{-1}f$ & $L^{-2}T^{-1}$\tabularnewline
Energy density & $E(p,\vec{r},t)=\frac{1}{c}\int d\Omega_{p}I(\vec{p},\vec{r},t)$ & $L^{-3}$\tabularnewline
Energy flux & $\vec{F}(p,\vec{r},t)=\int d\Omega_{p}\ \hat{p}I(\vec{p},\vec{r},t)$ & $L^{-2}T^{-1}$\tabularnewline
Momentum tensor & $P^{ij}(p,\vec{r},t)=\frac{1}{c}\int d\Omega_{p}\ \hat{p}^{i}\hat{p}^{j}I(\vec{p},\vec{r},t)$ & $L^{-3}$\tabularnewline
Higher moment tensors & $Q^{(n)}=\int d\Omega_{p}\ (\hat{p})^{n}I(\vec{p},\vec{r},t)$ & $L^{-2}T^{-1}$\tabularnewline
%\bottomrule
\end{tabular}
\end{table}
%\end{center}

\end{document}